\documentclass[preprint,aps,showkeys,showpacs]{revtex4}
\usepackage{graphicx}
\usepackage{dcolumn}
\usepackage{bm}

\preprint{}
\def\para{\par\noindent}
\def\sqr#1#2{{\vcenter{\vbox{\hrule height.#2pt
        \hbox{\vrule width.#2pt height#1pt \kern#1pt
          \vrule width.#2pt}
        \hrule height.#2pt}}}}

\newcount\notenumber

\def\note{\advance\notenumber by 1
\footnote{$^{\the\notenumber}$}} \baselineskip 20pt
\begin{document}
\title{Persistence in Random Bond Ising Models of a Socio-Econo Dynamics
 in High Dimensions}
 \author{S. \surname{Jain}}
\email{S.Jain@aston.ac.uk}
 \affiliation{Information Engineering, The Neural
Computing Research Group, School of Engineering and Applied Science,
Aston University, Birmingham B4 7ET, U.K.}
\author{T. \surname{Yamano}}
\affiliation{Social Science Research Institute, International
Christian University 3-10-2 Osawa, Mitaka 181-8585 Tokyo, Japan}
\date[]{Received October 17 2006}

\begin{abstract}

We study the persistence phenomenon in a socio-econo dynamics model
using computer simulations at a finite temperature on hypercubic
lattices in dimensions up to 5. The model includes a \lq social\rq\
local field which contains the magnetization at time $t$. The
nearest neighbour quenched interactions are drawn from a binary
distribution which is a function of the bond concentration, $p$. The
decay of the persistence probability in the model depends on both
the spatial dimension and $p$. We find no evidence of \lq
blocking\rq\ in this model. We also discuss the implications of our
results for applications in the social and economic fields.

\end{abstract}

\maketitle
\section{Introduction}
\para The persistence problem is concerned
 with the fraction of space
which persists in its initial $(t=0)$ state up to some later time
$t$. The problem has been extensively studied over the past decade
for pure spin systems at both zero [1-4] and non-zero [5]
temperatures.
\para Typically, in the non-equilibrium dynamics of spin systems
 at zero-temperature,
 the system is prepared initially in a random state and the fraction of
spins, $P(t)$, that persists in the same state as at $t=0$ up to some
later time $t$ is studied. For the pure ferromagnetic Ising model on
a square lattice the persistence probability has been found to decay
algebraically [1-4]
$$ P(t)\sim t^{-\theta},\eqno(1)$$
where $\theta\sim 0.22$ is the non-trivial persistence exponent
[1-3].
\para The actual value of $\theta$ depends on both the spin [6] and
spatial [3] dimensionalities; see Ray [7] for a recent review.
\para At non-zero temperatures [5], consideration of the global order
parameter leads to a value of $\theta_{global}\sim 0.5 $ for the pure
two-dimensional Ising model.
\para It has been only fairly recently established that systems
containing disorder [8-10] exhibit different persistence behaviour
to that of pure systems. A key finding [8-9,11] is the appearance of
\lq blocking\rq\ regardless of the amount of disorder present in the
system.
\para As well as theoretical models, the persistence phenomenon
has also been studied in a wide range of experimental systems and
the value of $\theta$ ranges from $0.19$ to $1.02$ [12-14],
depending on the system. A considerable amount of the recent
theoretical effort has gone into obtaining the numerical value of
$\theta$ for different models.

\para In this work we add to the knowledge and understanding regarding
persistence by presenting the initial results for the persistence
behaviour of a modified version of a recently proposed spin model
which appears to reproduce the intermittent behaviour seen in real
financial markets [15]. In the next section we discuss the model in
detail. In the subsequent section we give an outline of the method
used and the values of the various parameters employed. Section IV
describes the results and the consequent implications for using the
models in a financial or social context. Finally, in Section V there
is brief conclusion.
\section{The Modified Bornholdt Model}
\para The simulations were performed on a modified version of a spin
model of financial markets proposed recently by Bornholdt [15].

In the original Bornholdt model, $N$ market traders, denoted by Ising
spins $S_i(t), i=1\dots
N$, are located on the sites of a hypercubic lattice. The action of the
$i$th trader of buying or selling at time step $t$ corresponds to
the spin variable $S_i(t)$ assuming the value $+1$ or $-1$,
respectively. A local field, $h_i(t)$, determines the dynamics of
the spins. In particular,
$$h_i(t)=\sum_{<ij>}J_{ij}S_j(t)-\alpha
C_i(t)\sum_{j=1}^NS_j(t),\eqno(1)$$ where the first summation runs over
the nearest neighbours of $i$ only ($J_{ij}=J$, for nearest neighbours and
$J_{ij} =0$, otherwise), $\alpha
> 0$ couples to the magnetization, and $C_i(t)$ is a second spin
used to represent the strategy of agent $i$.

Subsequently, Yamano [16] worked with a model where the local field
is given by
$$h_i(t)=\sum_{<ij>}J_{ij}S_j(t)-\alpha\mid{\sum_{j=1}^NS_j(t)}\mid .\eqno(2)$$
Although here the strategy spin is omitted, the coupling constant is
retained. Furthermore, the nearest neighbour interactions are now
selected randomly, $J_{ij}=\pm J$. Each agent is updated according
to the following heat bath dynamics:
\[S_i(t+1)= \left\{
\begin{array}{c c}
+1 & \quad \mbox{with $q=[1+exp(-2\beta h_i(t))]^{-1}$,}\\
-1 & \quad \mbox{with $1-q$,}\\ \end{array} \right. \qquad\qquad\qquad\qquad\qquad\qquad\qquad\qquad (3)\] where $q$
is the probability of updating and $\beta$ is the inverse
temperature. In this model the return is defined in terms of the
logarithm of the absolute value of the magnetization,
$M(t)=\sum_{j=1}^NS_j(t)/N$, that is
$$ Return\ (t) = \ln\mid {M(t)}\mid -ln\mid {M(t-1)}\mid\eqno(4)$$
Simulations [16] in spatial dimensions ranging from $d=1$ to $d=5$
indicate that the modified version of the model reproduces the
required intermittent behaviour in the returns for suitable values
of the coupling constant and the temperature $T$; these are listed
in Table 1.
\para In this work we investigate the persistence behaviour of the
model where the local field is given by equation (2) but the nearest neighbour
interactions are selected from
$$P(J_{ij}) = (1-p)\delta (J_{ij} +1) + p\delta (J_{ij} -1),\eqno(5)$$.
where $p$ is the concentration of ferromagnetic bonds. Hence, we are
interested in determining the fraction of traders who have been at
time $t$ either buying or selling continuously since $t=0$. We will
also suggest a possible interpretation within sociophysics of the
model later on.

\section{Methodology}
\para
\begin{table}
\centering
\begin{tabular}{|c|c|c|c|}
  \hline
  {Dimension} & {$L$} & {$T_{int}$} \\ \hline
   1 & 4000001 & 3.5  \\ \hline
   2 & 2001 & 3.0 \\ \hline
   3 & 151 & 2.5 \\ \hline
   4 & 45 & 1.9  \\ \hline
   5 & 21 & 1.4  \\
  \hline
\end{tabular}
\caption{Values of the linear dimension $L$ of the lattices used in
the simulations. The coupling parameter $\alpha= 4.0$ in all cases.
Intermittent behaviour was observed in the returns when the
temperature was set at $T_{int}$ as given above.} \label{datavalues}
\end{table}

\para As mentioned in the previous section, for each spatial dimension $d$ we first
fine tune the temperature to reproduce intermittent behaviour in the
returns. As can be seen from Table 1, the temperature $T_{int}(d)$
decreases with $d$. For a given dimension, all subsequent
simulations are performed at that temperature. Averages over at
least 100 samples for each run were performed and the error-bars in
the following plots are smaller than the data points.

The value of each agent at $t=0$ is noted and the dynamics updated
according to equation (3).
 \para At each time step, we count the number of agents that still
persist in their initial $(t=0)$ state by evaluating
 $$n_i(t)=(S_i(t)S_i(0)+1)/2.\eqno(6)$$
 Initially, $n_i(0)=1$ for all $i$. It changes to zero when an agent
 changes from buying to selling or vice vera for the first time.
 Note that
 once $n_i(t) =0$, it remains so for all subsequent calculations.
\para The total number, $n(t)$,
of agents who have not changed their action by time $t$ is then
given by
$$ n(t)=\sum_{i}n_i(t).\eqno(7)$$
A fundamental quantity of interest is $P(t)$, the persistence
probability. In this problem we can identify $P(t)$ with the density
of non-changing agents [1].
$$P(t)=n(t)/N,\eqno(8)$$
where $N = L^d$ is the total number of agents present.
\section{Results}
\para  We now discuss our results.
In figure 1 we show a semi-log plot of the persistence probability
against time $t$ for a range of bond concentrations $0<p\le 0.5$ for
$d=1$. It\rq s clear from the plot that the data can be fitted to
$$P(t)\sim e^{-\gamma t},\eqno(9)$$
where we estimate $\gamma\sim 0.56$ from the linear fit.
\begin{figure}
\includegraphics{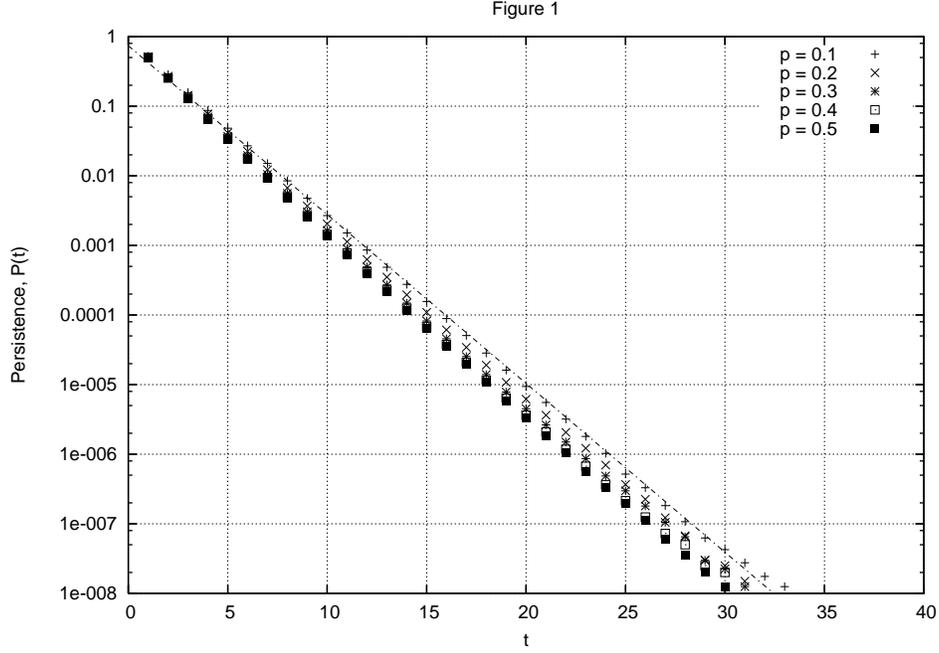}
\caption{Here we plot $\ln P(t)$ versus $t$ for $d=1$ over the range $0.1\le p\le 0.5$. The straight line, which
is a guide to the eye, has a slope of $-0,56$.} 
\end{figure}
\para
Figure 2 displays the results for $d=2$. Although once again there
is evidence for exponential decay, this time it would appear that
the value of the parameter $\gamma$ depends on the $p$. For $p=0.1$
we estimate $\gamma\sim 0.35$.
\begin{figure}
\includegraphics{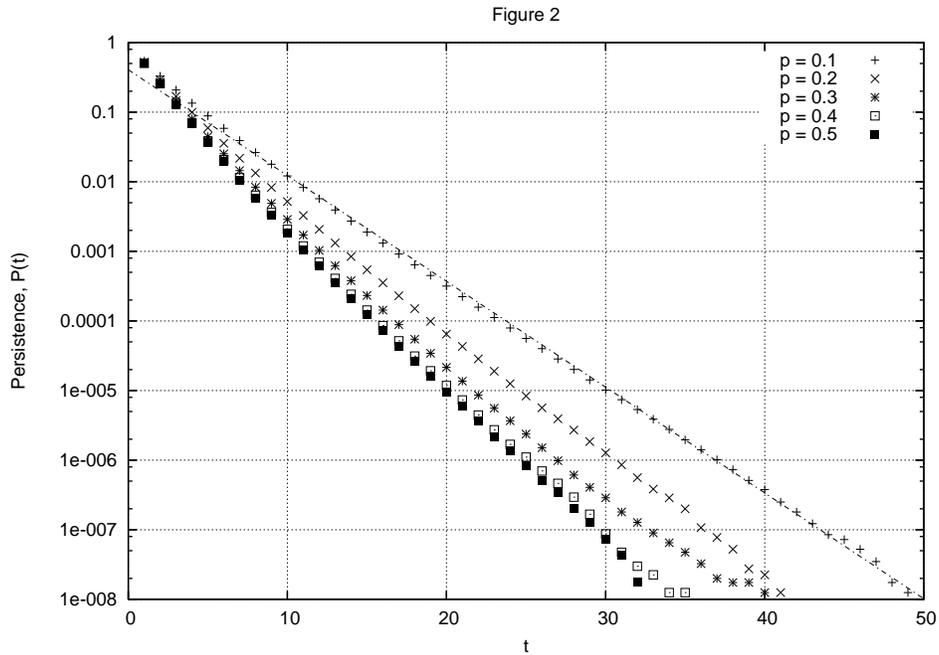}
\caption{A semi-log plot of the data for $d=2$. We see that here, in contrast to figure 1 for $d=1$, the slopes are
dependent on the bond concentrations. The linear fit shown is that for $p=0.1$ and the slope is $-0.35$.} 
\end{figure}
The results for the three-dimensional case are shown in figure 3.
Here we see clear evidence of the qualitative nature of the decay
depending on the bond concentration. For $p=0.5$ we have behaviour
very similar to the two cases considered earlier, namely exponential
decay. However, the decay is clearly non-exponential for $p=0.1$.
\begin{figure}
\includegraphics{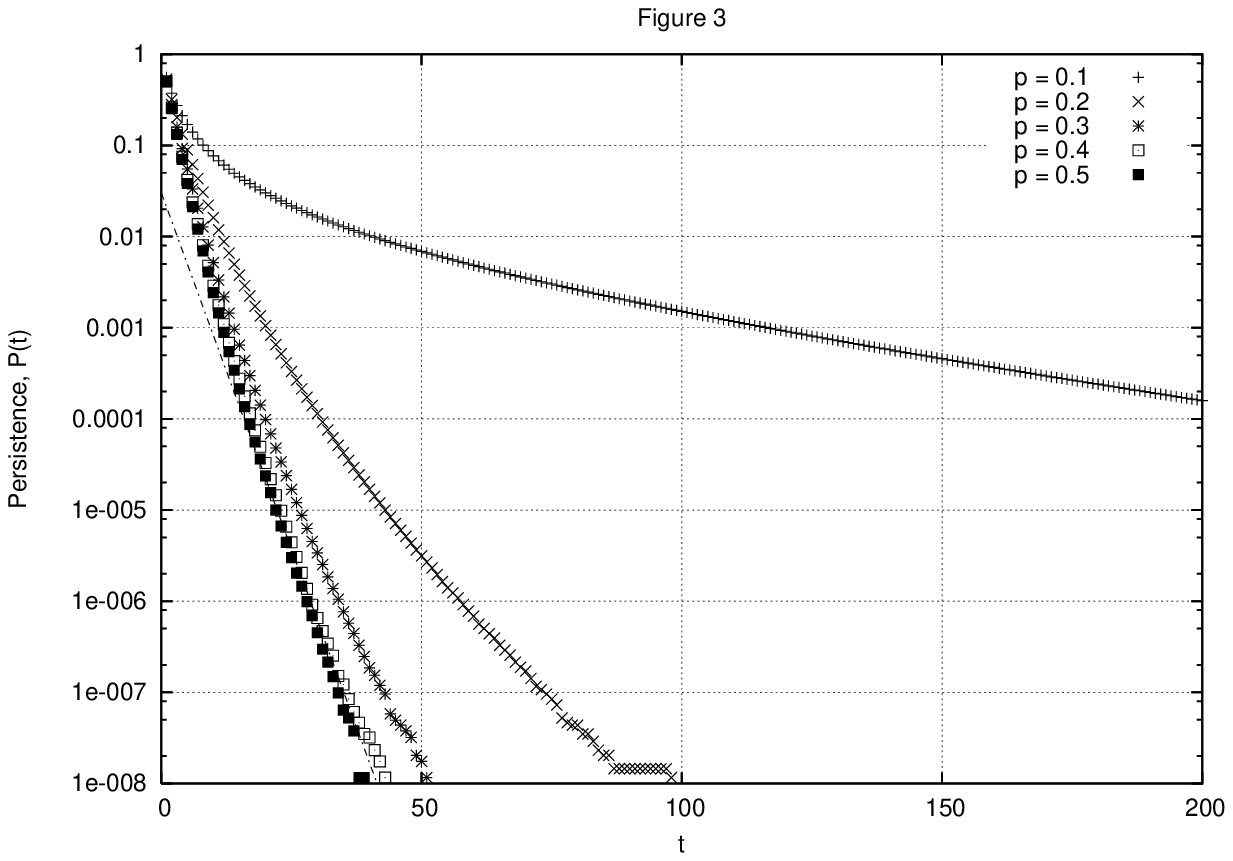}
\caption{A plot of $\ln P(t)$ against $t$ for $d=3$ for the same bond concentrations as earlier. The straight line, which is a 
guide to the eye, has a slope of $-0.36$ and indicates that the decay for $p=0.5$ is very similar to that found in lower dimensions.
The behaviour for $p=0.1$ is clearly non-exponential.} 
\end{figure}
The results in $d=4$ are very similar to those for $d=3$ and we will
not present them here. Instead, in figure 4 we show a log-log plot
of the persistence against time for $d=5$. The decay of $P(t)$ is
seen to be heavily dependent on the concentration of ferromagnetic
bonds. For low values of $p (\le 0.3$), we have a power-law decay at
long times as given by equation (1) with an estimated value of
$\theta\sim 0.5$. For higher value of $p$ the decay would appear not
to be a power-law but also not exponential in it\rq s nature.
\para
\begin{figure}
\includegraphics{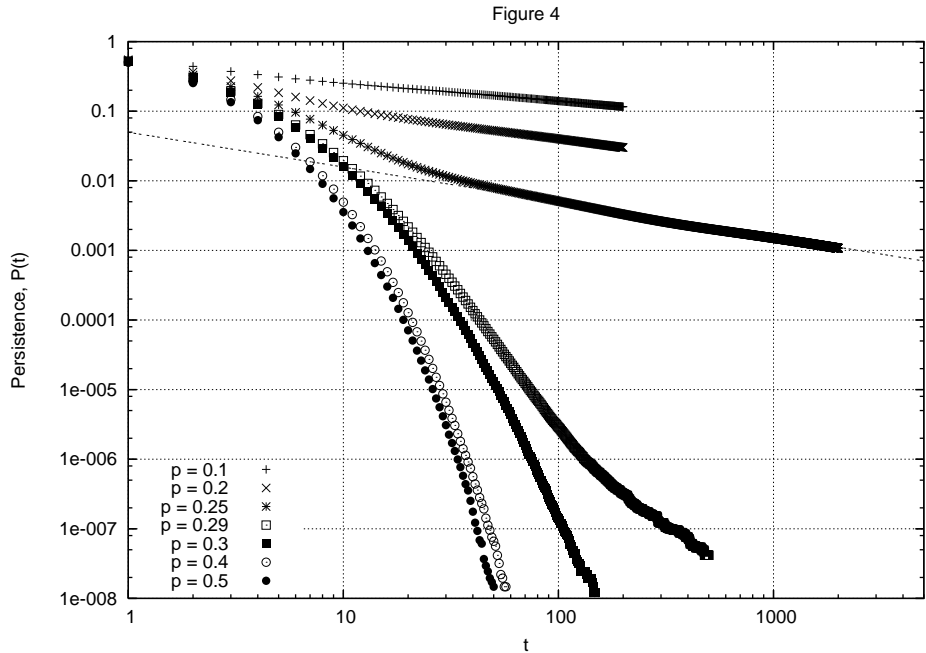}
\caption{Here we display the data for $d=5$ and selected bond concentrations as a log-log plot. Clearly the behaviour
depends crucially on the value of $p$. For low ($p\le 0.3$) values the decay is power-law. The straight line shown has a slope of
$-0.5$.} 
\end{figure}
\para
\section{Conclusion}
\para To conclude, we have presented the results of extensive simulations
for the persistence behaviour of agents in a model capturing some of
the features found in real financial markets. Although the model
contains bond disorder, we do not find any evidence of \lq
blocking\rq\ . The persistence behaviour appears to depend on both
the spatial dimensionality and the concentration of ferromagnetic
bonds. Generally, whereas in low dimensions the decay is
exponential, for higher dimensions and low values of $p$ we get
power-law behaviour.

The initial model was developed in an economic context. Power law
persistence in the case means the existence of traders who keep on
buying or selling for long durations. Furthermore, the presence of
\lq blocking\rq\ would be highly unrealistic for modelling the
dynamics because the traders would have access to a finite amount of
capital.

One can also interpret the model in a social context. Here the value
$S_i(t)=+1$ or $-1$ could represent an opinion. Here \lq
blocking\rq\ would be realistic and correspond to the proportion of
the population that is stubborn. Hence, any model exhibiting
exponential decay in the persistence probability would probably be
an unrealistic model.

Hence, we can use the behaviour of the persistence probability a
criterion to decide whether we have a realistic economic or social
model.

\begin{acknowledgments}
TY would like to thank Universitat Bremen where some of this work was performed.
TY also thanks L. Pichl for allowing him to use his CPU resources in the International Christian University Japan. The JSPS fellowship with the Grant-in-Aid from the Monbu-kagaku-sho is acknowledged for TY who also thanks L. Pichl for allowing him the use of CPU resources  in the International Christian University Japan. 
\end{acknowledgments}

\section*{References}
\begin{description}
\item {[1]} B. Derrida, A. J. Bray and C. Godreche, J. Phys. A:
Math Gen {\bf 27},
 L357 (1994).
\item {[2]} A. J. Bray, B. Derrida and C. Godreche, Europhys. Lett.
{\bf 27},
 177 (1994).
\item {[3]} D. Stauffer J. Phys. A: Math Gen {\bf 27}, 5029 (1994).
\item {[4]} B. Derrida, V. Hakim and V. Pasquier, Phys. Rev. Lett.
{\bf 75},
 751 (1995); J. Stat. Phys. {\bf 85}, 763 (1996).
 \item {[5]} S. N. Majumdar, A. J. Bray, S. J. Cornell, C. Sire,  Phys.
Rev. Lett. {\bf 77}, 3704 (1996).
 \item {[6]} B. Derrida, P. M. C. de Oliveira and D. Stauffer, Physica
{\bf 224A}, 604 (1996).
 \item {[7]} P. Ray, Phase Transitions {\bf 77} (5-7), 563 (2004).
\item {[8]} S. Jain, Phys. Rev. E{\bf 59}, R2493 (1999).
\item {[9]} S. Jain, Phys. Rev. E{\bf 60}, R2445 (1999).
\item {[10]} P. Sen and S. Dasgupta, J. Phys. A: Math Gen {\bf 37},
11949 (2004)
\item {[11]} S. Jain and H. Flynn, Phys. Rev. E{\bf 73}, R025701 (2006)
\item {[12]} B. Yurke, A. N. Pargellis, S. N. Majumdar
and C. Sire, Phys. Rev. E{\bf 56}, R40 (1997).
\item {[13]} W. Y.
Tam, R. Zeitak, K. Y. Szeto and J. Stavans, Phys. Rev. Lett. {\bf
78}, 1588 (1997).
\item {[14]} M. Marcos-Martin, D. Beysens, J-P
Bouchaud, C. Godreche and I. Yekutieli, Physica {\bf 214D}, 396
(1995).
\item {[15]} S. Bornholdt, Int. J. Mod. Phys. C{\bf 12}, 667 (2001).
\item {[16]} T. Yamano, Int. J. Mod. Phys. C{\bf 13}, 645 (2002).
\end{description}
\end{document}